\documentclass{aastex61}

\def\degpoint{\ifmmode ^{\rm{o}}\!. \else $^{\rm{o}}\!.$\fi}

\newcommand{\ms}{\mbox{m\,s$^{-1}$}}

\newcommand{\Mjup}{\mbox{$M_{\rm Jup}$}}

\usepackage{graphicx}
\usepackage{subfigure}
\usepackage{longtable}

\received{}
\revised{}
\accepted{\today}
\submitjournal{AJ}

\shorttitle{A New Orbital Solution For HD181433}
\shortauthors{Horner et al.}
\begin{document}

\title{The HD\,181433 Planetary System: Dynamics and a New Orbital Solution}

\correspondingauthor{Jonti Horner}
\email{jonti.horner@usq.edu.au}

\author[0000-0002-1160-7970]{Jonathan Horner}
\affiliation{Centre for Astrophysics, University of Southern Queensland, West Street, Toowoomba, QLD 4350, Australia}
\affiliation{Australian Centre for Astrobiology, UNSW Australia, Sydney, NSW 2052, Australia}

\author[0000-0001-9957-9304]{Robert A Wittenmyer}
\affiliation{Centre for Astrophysics, University of Southern Queensland, West Street, Toowoomba, 
QLD 4350, Australia}
\affiliation{Australian Centre for Astrobiology, UNSW Australia, Sydney, NSW 2052, Australia}

\author{Duncan J Wright}
\affiliation{Centre for Astrophysics, University of Southern Queensland, West Street, Toowoomba, QLD 4350, Australia}
\affiliation{School of Physics, UNSW Australia, Sydney, NSW 2052, Australia}
\affiliation{Australian Astronomical Observatory, PO Box 915, North Ryde, NSW 1670, Australia}

\author{Tobias C Hinse}
\affiliation{Chungnam National University, Department of Astronomy and Space Science, Daejeon 34134, Republic of Korea}

\author[0000-0001-6208-1801]{Jonathan P Marshall}
\affiliation{Academia Sinica, Institute of Astronomy and Astrophysics, 11F Astronomy-Mathematics Building, NTU/AS campus, No. 1,
Section 4, Roosevelt Rd., Taipei 10617, Taiwan}
\affiliation{Centre for Astrophysics, University of Southern Queensland, West Street, Toowoomba, 
QLD 4350, Australia}

\author{Stephen R Kane}
\affiliation{Department of Earth and Planetary Sciences, University of California, Riverside, CA 92521, USA}

\author{Jake T Clark}
\affiliation{Centre for Astrophysics, University of Southern Queensland, West Street, Toowoomba, QLD 4350, Australia}

\author{Matthew Mengel}
\affiliation{Centre for Astrophysics, University of Southern Queensland, West Street, Toowoomba, QLD 4350, Australia}

\author{Matthew T Agnew}
\affiliation{Centre for Astrophysics and Supercomputing, Swinburne University of Technology, Hawthorn, Victoria 3122, Australia}

\author[0000-0001-6311-8629]{Daniel Johns}
\affiliation{Department of Physical Sciences, Kutztown University, Kutztown, PA 19530, USA}


%% argument for \collaboration is the collaboration identifier. authors are

\begin{abstract}

We present a detailed analysis of the orbital stability of the HD\,181433 planetary system, finding it to exhibit strong dynamical instability across a wide range of orbital eccentricities, semi-major axes, and mutual inclinations. We also analyse the behaviour of an alternative system architecture, proposed by Campanella (2011), and find that it offers greater stability than the original solution, as a result of the planets being trapped in strong mutual resonance.

We take advantage of more recent observations to perform a full refit of the system, producing a new planetary solution. The best-fit orbit for HD~181433~d now places the planet at a semi-major axis of 6.60$\pm$0.22 au, with an eccentricity of 0.469$\pm$0.013. Extensive simulations of this new system architecture reveal it to be dynamically stable across a broad range of potential orbital parameter space, increasing our confidence that the new solution represents the ground truth of the system.

Our work highlights the advantage of performing dynamical simulations of candidate planetary systems in concert with the orbital fitting process, as well as supporting the continuing monitoring of radial velocity planet search targets.

\end{abstract}


\section{Introduction} \label{sec:intro}

Over the past 20 years, radial velocity surveys have discovered a plethora of multi-planet systems around nearby stars. These discoveries have revealed a diversity of orbital architectures, including compact systems \citep[e.g.][]{Lov11,TauCeti,Ang13,Fulton}, planets trapped in mutual mean-motion resonance \citep[e.g.][]{24Sex,HD155358,Laplace}, and others moving on startlingly eccentric orbits \citep[e.g.][]{ecc1,ecc2,ecc3}. Multi-planet systems enable detailed characterisation studies of orbital dynamics and the orbital evolution of systems as a function of time. Such dynamical studies may be used to constrain the physical properties of the system, such as the inclination of the planetary system relative to the plane of the sky \citep[e.g.][]{cor10,kan14}. 

More generally, dynamical studies are becoming a crucial component in the interpretation and determination of measured orbital properties. Such studies have revealed numerous cases of published solutions that place the proposed planets on orbits that prove dynamically unstable \citep[e.g.][]{JH11,RW13,hor14}, with some exhibiting catastrophic instability on timescales as short as a few thousand years \citep[e.g.][]{HWVir,QSVir}. Such extreme instability is a 'red flag' to the feasibility of a given exoplanetary system, and typically suggests that more observations are required in order to better constrain the published orbits. 

On other occasions, such studies have revealed that certain planetary systems are dynamically feasible, but only if the planets involved are trapped in mutual mean-motion resonance \citep[e.g.][]{HD204313,24Sex,73526,Tan13}. In cases such as these, the dynamical simulations afford an additional mechanism by which the range of plausible solutions for the system can be narrowed down. The widths of the 'stable' regions of parameter space afforded by such resonant interactions are typically smaller than the range of plausible solutions based solely on the observational data, allowing the dynamics to offer an important additional constraint on the architecture of the planetary system.

A published planetary system that is suspected of inherent instability can often be identified through visualisation of the orbits, or calculations of the Hill radii of the planets involved compared with minimum separations. One such example is the HD\,181433 system, first published by \citet{B09}. This intriguing system contains three planets, with masses of 0.02~$M_J$, 0.64~$M_J$ and 0.54~$M_J$. The orbital solution of \citet{B09} describes orbital periods ranging from 9 days to over 2000 days. 
However, the orbital eccentricities of outer two planets causes them to have the potential for significant close encounters, whereby the minimum separation of the orbits is 0.061 au, well within the Hill radius of each planet. The published orbital architecture thus quickly acquires an inherent instability that needs to be resolved. 

In this paper we present a new analysis of the HD\,181433 system that resolves the previous instability scenarios.  In Section~\ref{sec:System} we describe the HD\,181433 system in detail and provide stability simulations that demonstrate the system instability.  In Section~\ref{sec: Dynamics} we provide a description of dynamical simulations used to study exoplanet orbital stability.  Section~\ref{sec: SimResults} gives the results of our dynamical stability tests for the two published solutions for the HD\,181433 system. In section~\ref{sec:NewStuff} we present an analysis of new radial velocities that more than double the previously published radial velocity time series, and use these to provide a new orbital solution %including the exploration of a possible fourth planet. 
In Section~\ref{sec:NewDyn} we show the results of a detailed stability simulation based upon our revised orbital solution and demonstrate that the new orbital architecture exhibits long-term stability. We provide discussion and concluding remarks in Section~\ref{sec:Discussion}.

\section{The HD\,181433 System} \label{sec:System}

HD\,181433 is an old ($\sim$ 6.7 Gyr) high-metallicity star ([Fe/H] = $0.41 \pm 0.04$), located at a distance of 26.76 parsecs \citep{VLee07,Trev11}. It is cooler, less luminous, and rotates more slowly than the Sun. In \cite{B09}, its spectral classification is erroneously given as K3 IV, which would make it a sub-giant - a classification that is strongly at odds with its relatively low luminosity ($\sim$ 0.308 $L_\odot$). By contrast, the updated catalogue of stellar parameters detailed in \cite{VLee07} gives a spectral class of K5V for the star, which is in far better agreement with the published luminosity.

On the basis of 107 radial velocity measurements of HD\,181433 obtained over a period of four years using the HARPS spectrograph on the 3.6-m ESO telescope at La Silla, Chile, \cite{B09} announced the discovery of three planets orbiting the star. The proposed planetary system \citep[][see Table~\ref{table:plt_sol_Bouchy}]{B09} features three planets: a hot super-Earth, with an orbital period of 9.4 days, and two giant planets (of mass 0.64 and 0.54 times that of Jupiter) moving on orbits with periods of 2.6 and 6 years, respectively. In stark contrast to our own Solar system, the orbital eccentricities for the three planets are all relatively high: 0.396 $\pm$ 0.062, 0.28 $\pm$ 0.02 and 0.48 $\pm$ 0.05, respectively. The innermost planet is sufficiently far from the outer two that it is reasonable to assume that its orbit is not strongly perturbed by their presence. However, it is concerning to note that the nominal best-fit orbits for the two outer planets cross one another \citep{B09}, with the outermost planet (HD\,181433\,d) having a periastron distance of 1.56 au, well inside the semi-major axis of the orbit of HD\,181433\,c (1.76 au).

%Model solutions of previous works (Bouchy & Camponella)

\begin{table}[b]
\centering
\caption{Stellar parameters for HD~181433. For those parameters for which two values are presented, the first is that used in \cite{B09}, whilst the second is a more recent, updated value. In the case of the parallax and distance values, the third value given is that taken from the latest \textit{Gaia} data release - the most up-to-date values available. [1] The absolute V-band magnitude was calculated using the apparent V-band magnitude (obtained from the Simbad database) and the distance \citep{VLee07}, assuming no interstellar extinction. [2] We note that \cite{Trev11} provide a different mass for HD\,181433 to that used in \cite{B09}. In order that our results are directly comparable to those of \cite{B09}, we use the older value of the mass in our fitting process.}
\label{table:StarParams}
\begin{tabular}{ccc}
\hline \hline
Parameter & Value & Reference \\
\hline\hline
Spectral type &  K3 IV & Hipparcos, via \cite{B09} \\
                      &  K5V & \cite{VLee07} \\ \hline
Age [Gyr] & 6.7 $\pm$ 1.8 & \cite{Trev11} \\ \hline
Parallax [mas]   & 38.24 & Hipparcos, via \cite{B09} \\
                          & 37.37 $\pm$ 1.13 & \cite{VLee07} \\
                          & 37.17871 $\pm$ 0.03089 & \cite{Gaia2}\\ \hline
Distance [pc]  & 26.15 & Hipparcos, via \cite{B09} \\
                       & 26.76 & \cite{VLee07} \\ 
                       & 26.89711 $\pm$ 0.02232 & \cite{Gaia} \\ \hline
$m_v$   & 8.4   & Hipparcos, via \cite{B09} \\
             & 8.38 & \cite{Simbad} \\ \hline
$M_v$    & 6.31 & Hipparcos, via \cite{B09} \\
              & 6.24 & [1] \\ \hline
$B - V$  & 1.01 & Hipparcos, via \cite{B09} \\
              & 1.006 & \cite{VLee07} \\ \hline
Luminosity [$L_\odot$] & 0.308 $\pm$ 0.026 & \cite{Sousa08} \\ \hline
Mass [$M_\odot$] & 0.78 $M_\odot$ & \cite{Sousa08} \\
                             & 0.86 $M_\odot$ $\pm$ 0.06 & \cite{Trev11} [2] \\ \hline
$T_{eff}$ [K] & 4962 $\pm$ 134 & \cite{Sousa08} \\
                     & 4902 $\pm$ 41 & \cite{Trev11} \\ \hline
log$g$ & 4.37 $\pm$ 0.26     &\cite{Sousa08} \\
	& 4.57 $\pm$ 0.04 & \cite {Trev11} \\ \hline
[Fe.H] & 0.33 $\pm$ 0.13 & \cite{Sousa08} \\
           & 0.41 $\pm$ 0.04 & \cite{Trev11} \\  \hline
$v$sin$i$ [km s$^{-1}$] & 1.5 & \cite{B09} \\ \hline
log$R^{'}_{HK}$ & -5.11 & \cite{B09} \\ \hline
$P_{rot}$ [days] & 54 & \cite{B09} \\ \hline
\hline
\end{tabular}
\end{table}

\begin{table}[b]
\centering
\caption{Orbits and physical parameters of HD~181433's planets according to \cite{B09} (their Table 3). \label{table:plt_sol_Bouchy}}
\begin{tabular}{lccc}
\hline
Parameter & HD~181433b & HD~181433c & HD~181433d \\
\hline\hline
$P$ (d) & 9.3743~$\pm$~0.0019 & 962~$\pm$~15 & 2172~$\pm$~158 \\
$T_{\rm peri}$ (BJD-2400000) &  54 542.0~$\pm$~0.26 & 53 235.0~$\pm$~7.3 & 52 154~$\pm$~194 \\
$e$ & 0.396~$\pm$~0.062 & 0.28~$\pm$~0.02 & 0.48~$\pm$~0.05 \\
$\omega$ (\degr) & 202~$\pm$~10 & 21.4~$\pm$~3.2 & 330~$\pm$~13 \\
$V$ (km/s) & \multicolumn{3}{c}{40.2125~$\pm$~0.0004} \\
$K$ (m/s) & 2.94~$\pm$~0.23 & 16.2~$\pm$~0.4 & 11.3~$\pm$~0.9 \\
$m \sin i$ ($M_{\rm Jup}$) & 0.024 & 0.64 & 0.54 \\
$a$ (au) &  0.080 & 1.76 & 3 \\
\hline
\end{tabular}
\end{table}

Mutually crossing orbits are almost always dynamically unstable, unless 
the objects involved are protected from close encounters by the 
influence of mean-motion resonances - as is seen in our own Solar system 
for the Jovian and Neptunian Trojans \citep[e.g.][]{NeptCent,QR322,LC18}, and the 
Plutinos \citep[e.g.][]{Plut1,Plut2}. Indeed, \citet{C11} noted that the 
\citet{B09} solution for HD\,181433 was dynamically unstable, and 
reanalysed the original radial velocity data in search of a stable 
solution in the neighbourhood of the formal best fit. They found that 
the system would be stable if the two giant planets were locked in a 
5:2 mean-motion resonance, which would protect them from mutual close 
encounters.  The stable best-fit solution found by \citet{C11} had a 
slightly worse $\chi^2$ than the (unstable) statistical best fit (as 
shown in Table~\ref{table:plt_sol_Campanella}).  Such instances are not 
unusual, as other strongly interacting systems have been shown to fall 
into stable configurations if allowed some flexibility around the 
statistical best fit \citep[e.g.][]{trifonov14, 30177}.

\begin{table}
\centering
\caption{Orbits and physical parameters of HD~181433's planets according to \cite{C11} (their Table 1). \label{table:plt_sol_Campanella}}
\begin{tabular}{lccc}
\hline
Parameter & HD~181433b & HD~181433c & HD~181433d \\
\hline\hline
$P$ (d) & 9.37459 & 975.41 & 2468.46 \\
$T_{\rm peri}$ (BJD-2400000) & 7788.9185 & 7255.6235 & 6844.4714 \\
$e$ & 0.38840 & 0.26912 & 0.46626 \\
$\omega$ (\degr) & 202.039 & 22.221 & 319.129 \\
$V$ (km/s) & \multicolumn{3}{c}{40.212846~$\pm$~0.00136} \\
$K$ (m/s) &  2.57 & 14.63 & 9.41 \\
$m \sin i$ ($M_{\rm Jup}$) & 0.02335 & 0.65282 & 0.52514 \\
$a$ (au) & 0.08013 & 1.77310 & 3.29347 \\
\hline
\end{tabular}
\end{table}

\section{Dynamical Simulations of Exoplanet Systems} \label{sec: Dynamics}

In a number of previous works, we have examined the dynamical stability of proposed exoplanetary systems, in order to provide a 'sanity check' as to their veracity \citep[see e.g.][]{HR8799,142paper,HD204313}. In some cases, those simulations reveal that the planets as proposed are not dynamically feasible \citep[e.g.][]{JH11,HWVir,QSVir,hor14,HUAqr2}, suggesting that further observations are needed to refine their orbits. In other cases, our simulations allow the orbits of proposed planets to be better constrained, revealing them to only be stable if they are trapped in mutual mean motion resonance \citep[e.g.][]{HD155358,24Sex,73526}.

In the process, we have developed a standard method for analysing such systems, creating dynamical maps that show the context of the orbital solutions proposed. Using the {\it n}-body dynamics package {\sc Mercury} \citep{Mercury}, we run a large number (typically 126,075) of individual realisations of the planetary system in question, placing the planet with the least constrained orbit on a different initial orbit each time. In each of those simulations, we follow the evolution of the planets in question for a period of 100 million years, or until they either collide with one another, are ejected from the system, or collide with the central body.

In the case of the HD\,181433 system, as discussed above, the innermost planet (with the $\sim$9 day orbital period) is so distant from the others that it is almost certainly totally decoupled from their dynamical influence. For that reason, in the simulations that follow, we add the mass of that planet to that of the central star, and do not integrate its orbital evolution\footnote{Including the evolution of the innermost planet would require the use of an unfeasibly short integration timestep, as well as the calculation of several post-Newtonian terms. By including this planet with the central mass, our simulations can run in a reasonable amount of time, and can focus on the behaviour of the two planets that are of dynamical interest in this work.}

For both the \citet{B09} and \citet{C11} solutions, we carried out a highly detailed suite of primary integrations. For these simulations, we held the initial orbit of HD\,181433\,c fixed at its nominal best fit value, and incrementally varied the initial orbit of HD\,181433\,d around the best-fit solution proposed. In each case, we tested 41 different values for the semi-major axis of that planet, distributed evenly across the full $\pm 3\sigma$ range detailed in Tables~\ref{table:plt_sol_Bouchy} and \ref{table:plt_sol_Campanella}\footnote{We note that \cite{B09} provided no estimate of the uncertainty of the semi-major axes of the orbits of the planets, whilst \cite{C11} gave a single solution with no uncertainties. As such, we use an uncertainty for the semi-major axis for the \cite{B09} calculated directly from the uncertainty in its orbital period (which yields $\pm 0.155$ au). We then apply the \cite{B09} uncertainties directly to the \cite{C11} solution to generate the clones for that scenario.}. At each of those semi-major axes, we tested 41 unique values of eccentricity, again evenly distributed across the full $\pm 3\sigma$ range detailed above. For each of those locations in $a-e$ space, we tested 15 unique values of $\omega$, with five unique values of mean-anomaly tested for each unique $\omega$ examined. This gave a grid of 41 $\times$ 41 $\times$ 15 $\times$ 5 = 126,075 simulations, which were performed with an integration time-step of 40 days using the Hybrid integration package within \textsc{Mercury}.

To complement these $n$-body simulations, we produced a MEGNO \citep[Mean Exponential Growth factor of Nearby Orbits;][]{Cin00,GZ01,Cin03} map of the $a-e$ space around the best-fit solution for the orbit of HD\,181433\,d, following our earlier work \citep[e.g.][]{SWLyn,Bruna,Wood17}. This map was created with a resolution of 720 x 640, with a single test particle being integrated forwards in time for five thousand years for each pixel in the phase-space, using the Gragg-Bulirsch-Stoer method \citep{Hair}. 

The resulting MEGNO map shows the chaoticity of the region of $a-e$ phase space around the best-fit orbit for HD\,181433\,d, categorised at each point in terms of a parameter $< Y >$, which is proportional to the Lyapunov characteristic exponent, which characterises the rate at which a given two orbits will diverge. For more details on this process, we direct the interested reader to \citet{Cin00}, \citet{GZ01}, \citet{Cin03}, \citet{GC04} and \citet{Hinse10}. 

Orbits that display quasi-periodic behaviour, or are typically dynamically stable, will yield values for $<Y>$ of approximately 2.0. By contrast, for chaotic orbits, the value of $<Y>$ will diverge from 2.0 rapidly as time passes. As a result, mapping the value of $<Y>$ as a function of initial orbital parameters provides an independent means of quantifying the stability or chaoticity of a given scenario.

In addition, to investigate the impact of the mutual inclination of the two planets in question, we built on our earlier work \citep{JH11,QSVir,hor14}, and carried out subsidiary integrations of the \cite{B09} solution that each covered 11,025 unique solutions (21 $\times$ 21 $\times$ 5 $\times$ 5 in $a-e-\omega-M$). Five such simulations suites were carried out, considering mutual inclinations between the two planets of 5$^\circ$, 15$^\circ$, 45$^\circ$, 135$^\circ$ and 180$^\circ$. Our results are presented below.

\section{The Stability of the Bouchy and Campanella solutions} 
\label{sec: SimResults}

\begin{figure}
\plotone{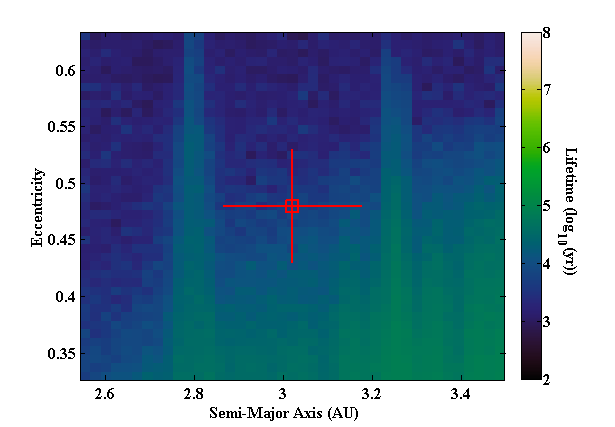}
\caption{The stability of the orbit of HD\,181433\,d, for the orbital solution proposed by \citet{B09}, as a function of the planet's orbital semi-major axis (a) and eccentricity (e). The location of the nominal best-fit orbit is marked by the hollow square, with the solid red lines radiating from that point showing the $\pm$1$\sigma$ errors on those values. Each coloured grid point shows the mean lifetime of 75 distinct dynamical simulations, testing a variety of plausible values for the planet's longitude of periastron ($\omega$) and mean anomaly ($M$). The nominal best-fit orbit, and the region bounded by the $\pm$1$\sigma$ errors, falls in an area of significant dynamical instability, featuring mean lifetimes of order 10,000 years.}
\label{OriginalOrbits}
\end{figure}

%----------------------------------------------

\begin{figure}
\includegraphics[width=1.0\textwidth]{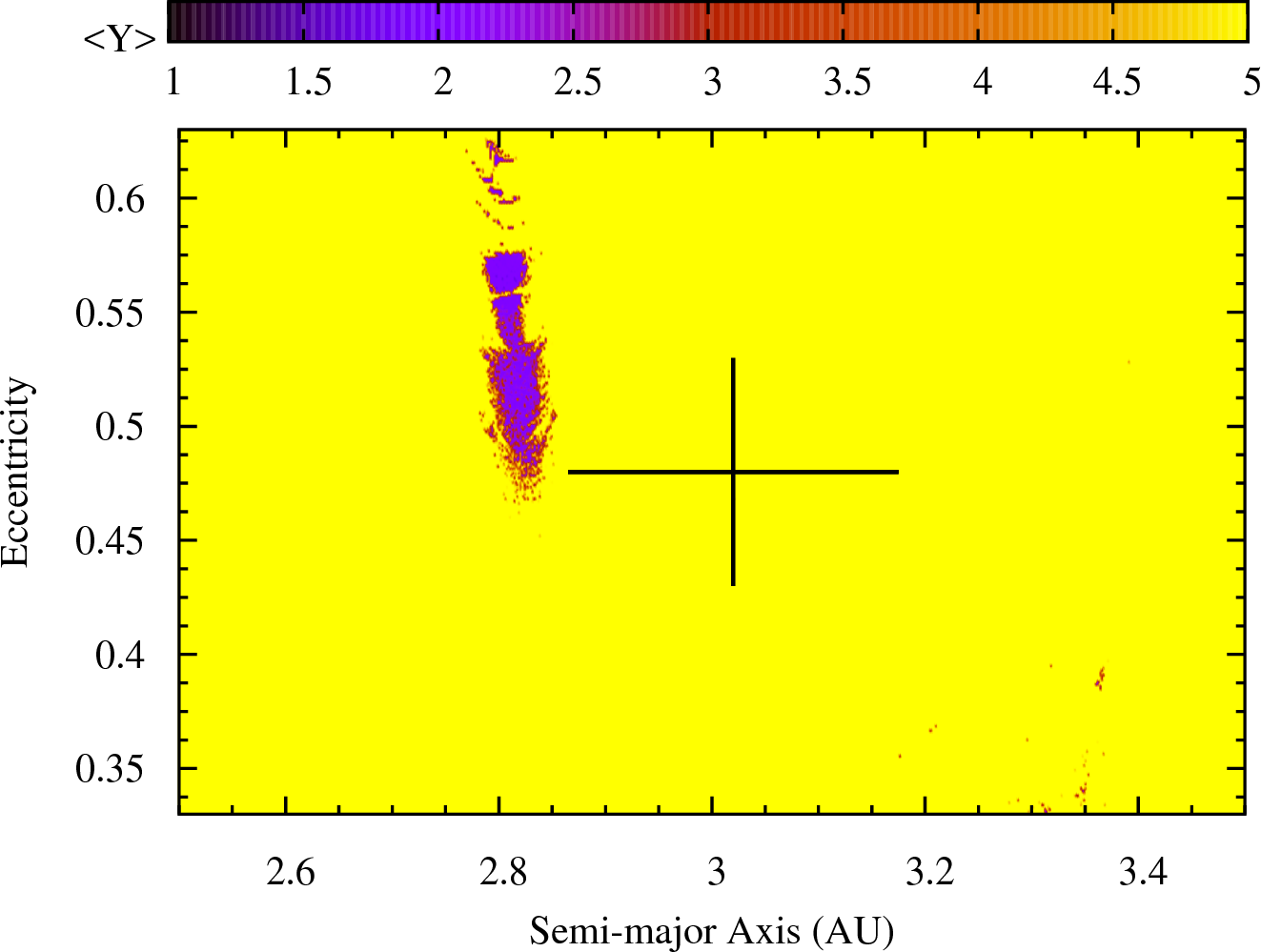}

\caption{MEGNO map of the $a-e$ space around the best-fit solution for the orbit of HD\,181433\,d, as proposed by \cite{B09}. Here, the colour indicates the chaoticity of the system for each given initial condition - with low values indicating stability, and high values pointing to a highly unstable orbit. As seen in Figure~\ref{OriginalOrbits}, the solution proposed by \cite{B09} lies in a region of extreme dynamical instability, and the proposed planetary solution is clearly not dynamically feasible.}

\label{bouchymegno}
\end{figure}

%----------------------------------------------

In Figure~\ref{OriginalOrbits}, we present the results of our simulations of the two outermost planets in the \citet{B09} solution for the HD\,181433 system. Across the range of orbits allowed within $\pm$3$\sigma$ of the nominal best-fit values, the stability of the system varies by around three orders of magnitude. Even the most stable solutions, however, are dynamically unstable on timescales of less than one million years. These findings are supported by the MEGNO mapof that region of $a-e$ space, which can be seen in Figure~\ref{bouchymegno}. The whole region around the best-fit orbit is a sea of extreme chaoticity. Our results therefore suggest that the system, as proposed in the discovery work, is not dynamically feasible. 

\begin{figure}
\gridline{\fig{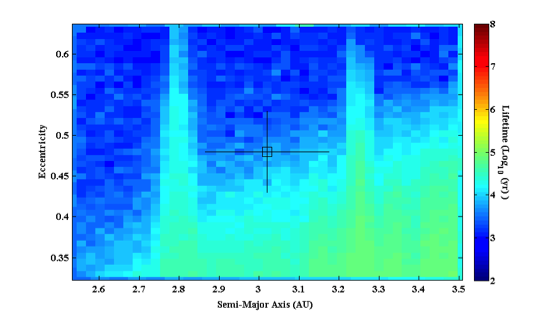}{0.5\textwidth}{Coplanar}
          \fig{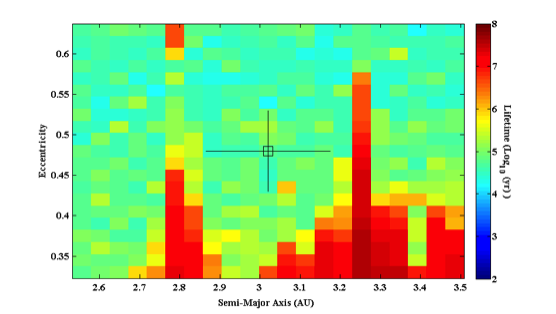}{0.5\textwidth}{5$^\circ$}}
\gridline{\fig{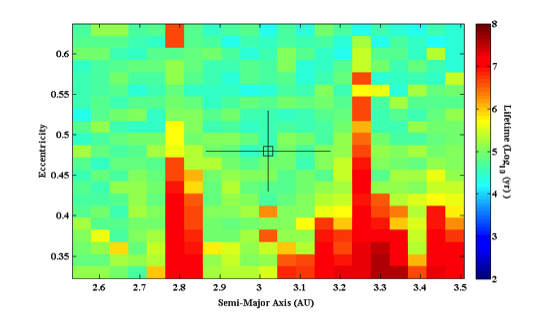}{0.5\textwidth}{15$^\circ$}
          \fig{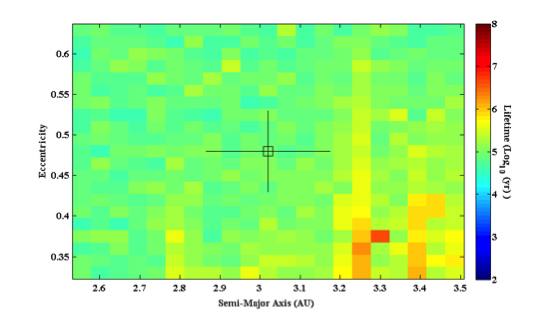}{0.5\textwidth}{45$^\circ$}}
\gridline{\fig{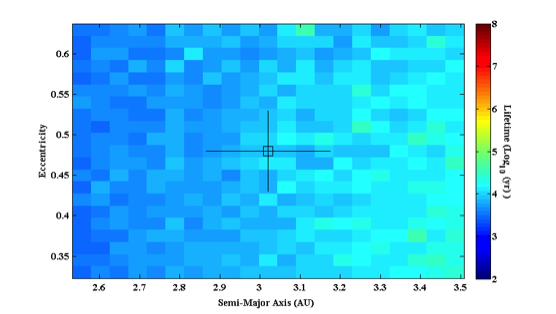}{0.5\textwidth}{135$^\circ$}
          \fig{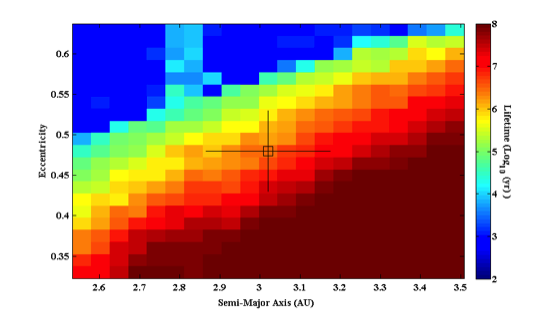}{0.5\textwidth}{180$^\circ$}}
\caption{Dynamical stability of the HD\,181433 planetary system, as proposed by \citet{B09}, as a function of the mutual inclination between the orbits of HD\,181433\,c and HD\,181433\,d. The left hand column shows the stability of scenarios where the two planets have a mutual inclination of 0$^\circ$ (coplanar case, top left), 5$^\circ$ (centre left) and 15$^\circ$(lower left). The right hand column shows the stability of scenarios where the planets have mutual inclinations of 45$^\circ$ (top), 135$^\circ$ (middle) and 180$^\circ$ (bottom). For clarity, the colour scale is the same in all figures, stretching from a mean lifetime of 10$^2$ years (dark blue) to 10$^8$ years (dark red). Interestingly, modest inclinations (5$^\circ$ and 15$^\circ$) offer significantly improved prospects for dynamical stability over the coplanar case. }
\label{Inc}
\end{figure}

In Figure~\ref{Inc}, we present the results of our simulations investigating the influence of the mutual inclination between the orbits of HD\,181433\,c and HD\,181433\,d, for the \citet{B09} solution. Interestingly, it is immediately apparent that a moderate mutual inclination (5$^\circ$ or 15$^\circ$, middle and lower panels on the left hand side, respectively) results in narrow strips of enhanced stability that stretch up to relatively high eccentricities. These regions of enhanced stability, around semi-major axes of 2.8 and 3.25 au, are the result of mutual mean-motion resonances between the two planets; the 5:2 mean-motion resonance \citep[as discussed in][]{C11} lies at around 3.24 au, while the 2:1 mean-motion resonance lies at 2.79 au. Both features are also visible in the results for the coplanar and 45$^\circ$ integrations, though in neither case do they offer sufficient enhancements to the system's stability that the planets might reasonably be expected to survive on timescales comparable to the lifetime of the star. If the planets are placed on orbits inclined at 180$^\circ$ to one another, then a large region of dynamical stability can be seen, with the nominal best-fit orbit for HD\,181433\,d lying on the boundary between stable and unstable solutions. This result is not unexpected - such retrograde solutions are almost always highly stable unless they feature mutually crossing orbits \citep[e.g.][]{EC10,JH11,NNSer,RW13,songhu,ramm16}.
  
\begin{figure}
\plotone{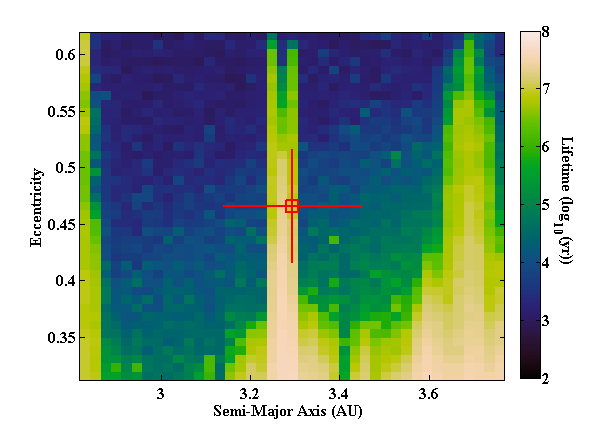}
\caption{The stability of the orbit of HD\,181433\,d as a function of that planet's orbital semi-major axis (a) and eccentricity (e) for the solutions presented in \citet{C11}. Once again, the location of the nominal best-fit orbit is marked by the hollow square, with the solid red lines radiating from that point showing the $\pm$1$\sigma$ errors on those values (taken from \citet{B09}, since the \citet{C11} solution was presented with no uncertainties). Each coloured grid point shows the mean lifetime of a total of 75 distinct dynamical simulations, testing a variety of plausible values for the planet's longitude of periastron and mean anomaly. In stark contrast to the orbital stability of the \citet{B09} solution, the \citet{C11} solution lies in the middle of a narrow region of dynamical stability resulting from the two planets being trapped in mutual 5:2 MMR. }
\label{C11Orbits}
\end{figure}

In Figure~\ref{C11Orbits}, we present the results of our simulations of the \citet{C11} solution for the HD\,181433 planetary system. Since no uncertainties are given in \citet{C11}, we chose to use the uncertainties from \citet{B09} as the basis for our integrations. This meant that we tested a wide variety of potential orbital architectures distributed evenly around the best-fit case presented in \citet{C11}, and that our results can be directly compared to those for the integrations carried out to study the \citet{B09} solution. Our results are presented in Figure~\ref{C11Orbits}.

It is immediately apparent that the solution presented in \citet{C11} exhibits significantly greater dynamical stability than that put forth in \citet{B09}. The best-fit orbit for HD\,181433\,d is now located noticeably further from the central star, placing it in 5:2 MMR with HD\,181433\,c.

The same broad features can be seen in the dynamical maps for the two solutions, but the regions of stability offered by the 2:1 and 5:2 MMRs (around $\sim$2.8 au and 3.25 au, respectively, in both cases) are significantly more stable in the case of the \citet{C11} solution than was the case for the \citet{B09} solution. An additional region of dynamical stability located at around 3.7 au is the result of the 3:1 MMR between the two planets. The difference in stability within the resonant regions between the two models is the direct result of differences in the initial mean anomaly and argument of periastron for the two planets in the two models. By targeting the solution with the best dynamical stability, Campanella's solution places the two planets on orbits that, when resonant, are protected from close encounters by the influence of the resonance (in much the same way that the Plutinos in our own Solar system are protected from close encounters with Neptune by the influence of their 2:3 MMR). With different initial angular values, the \citet{B09} solution results in a significant fraction of the resonant scenarios experiencing catastrophic close encounters between the two planets on remarkably short timescales.

\begin{figure}
\gridline{\fig{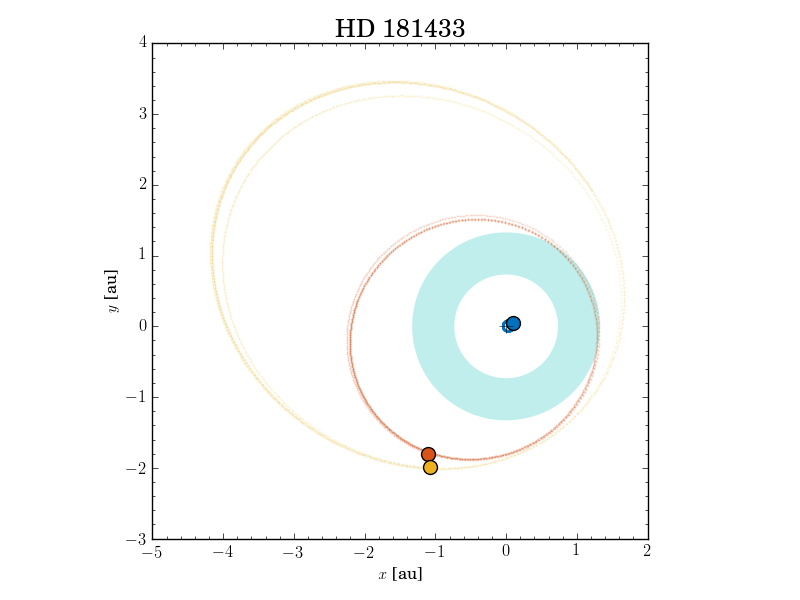}{0.49\textwidth}{}
          \fig{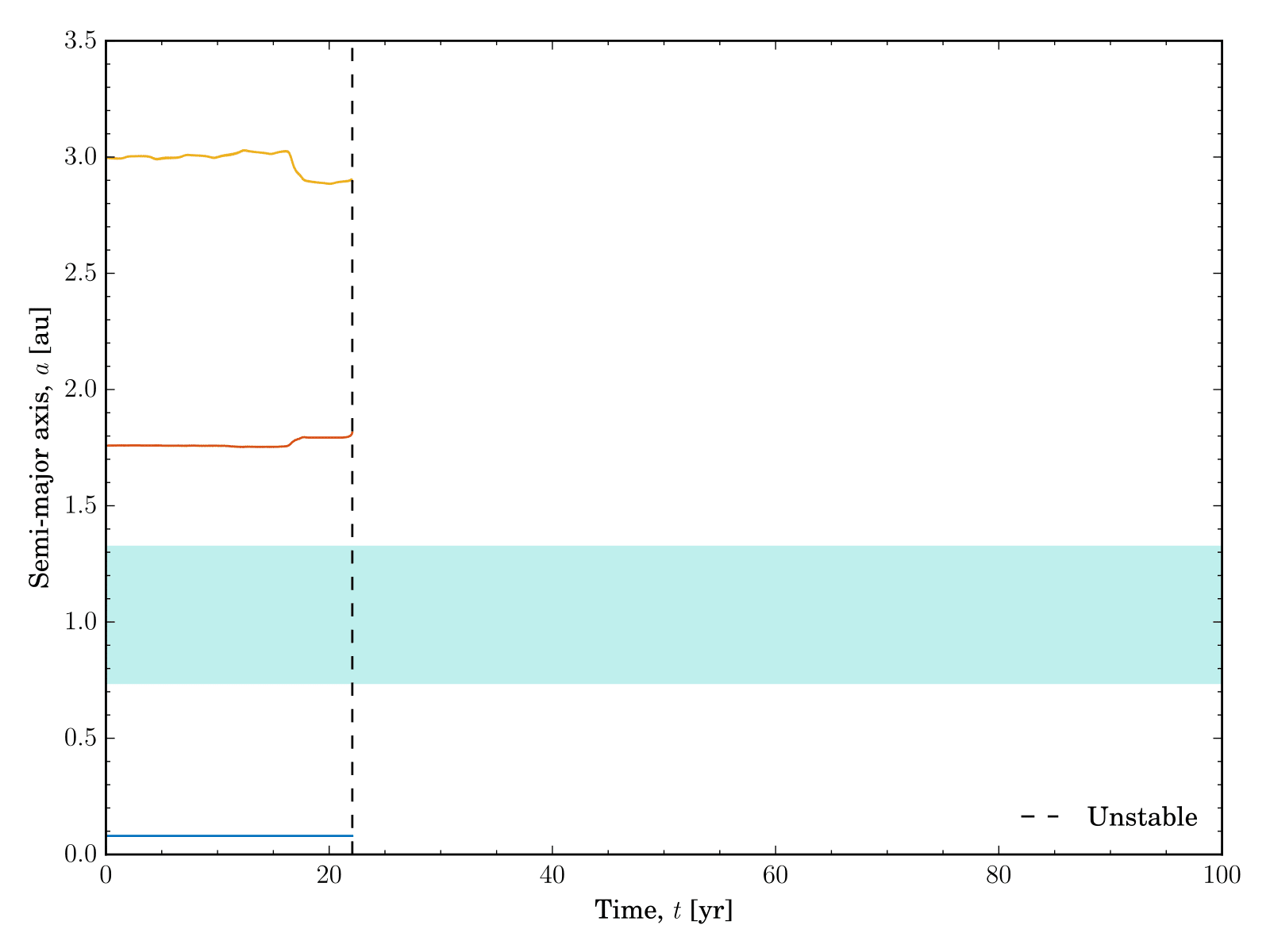}{0.49\textwidth}{}}
\gridline{\fig{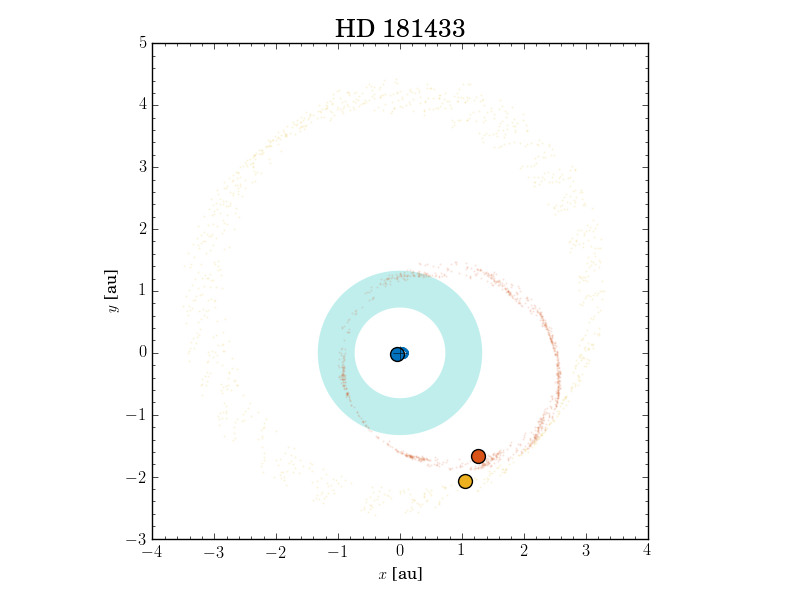}{0.49\textwidth}{}
          \fig{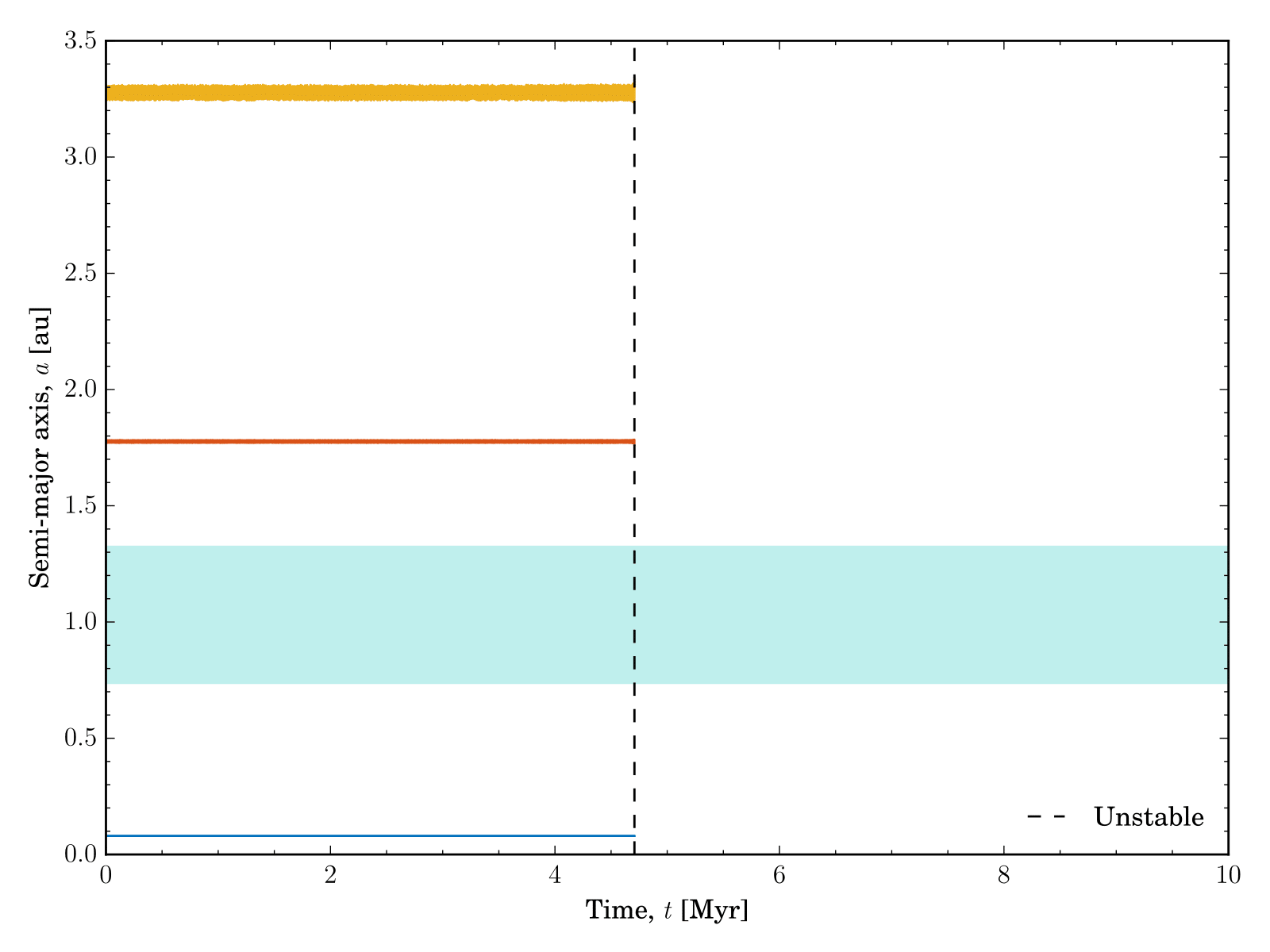}{0.49\textwidth}{}}
\caption{The evolution of the best-fit solutions for the orbits of the three planets in the HD\,181433 system, as proposed in \cite{B09} (top) and \cite{C11} (bottom). In both cases the orbital evolution of the planets was followed until they experienced their first close encounter within one Hill radius. In the case of the \cite{B09} solution, this occurred after just over 20 years, whilst the \cite{C11} solution remained stable until 4.7 million years had elapsed. The left-hand plots present a cartesian representation of the evolution, with the red and yellow filled circles showing the location of the two outer planets at the time of that first close encounter. The right-hand plots show the evolution of the semi-major axes of the planets as a function of time - with the vertical dashed line marked the point at which the first close encounter occurred.}
\label{unstable}
\end{figure}

To further illustrate the degree of instability offered by the orbital solutions proposed by \cite{B09} and \cite{C11}, we integrated the best-fit solutions proposed in those works forward in time using the SWIFT $N$-body software package \citep{Levison1994}, specifically the Regularised Mixed Variable Symplectic (RMVS) method, until either of the two outer planets approached within one Hill radius of the other. Both simulations used a time step equal to 1/50th the orbital period of the innermost body. In the case of the \cite{B09} solution, that first very close encounter occurred after just 20 years, whilst for the \cite{C11} solution, the first encounter within one Hill radius occurred after 4.7 million years. To illustrate this extreme instability we present the results of these simulations in Figure~\ref{unstable}, which shows the full evolution of the system in both cases until that first deep close encounter.

%----------------------------------------------

%----------------------------------------------

\section{A New Solution} \label{sec:NewStuff}

Since \citet{B09} published their three-planet solution for the HD\,181433 system, a large number of additional radial velocity measurements have obtained, and the spectra are publicly available in the ESO archive. Since it is well established that radial velocity fitting processes often initially exaggerate the eccentricity of planetary orbits \citep[e.g.][]{shen08,SOJ09,songhu}, and that new data can often yield dramatically different solutions for a given system, we felt that it would be prudent to obtain a new solution for the system, based on the new data. To obtain the longest possible time series, we obtained the publicly available HARPS spectra from the ESO archive and extracted the DRS radial velocities to perform a new analysis on a total of 200 observational epochs. 

\subsection{Recovering the inner planets}

We approach the fitting process in a traditional manner, by successive removal of Keplerian orbits based on their signals in the generalised Lomb-Scargle periodogram \citep{lomb76, scargle82, zk09}. We then use the \textit{Systemic Console} version 2.2000 \citep{mes09} to perform the orbit fitting and uncertainty analysis.  

\begin{figure}
\gridline{\fig{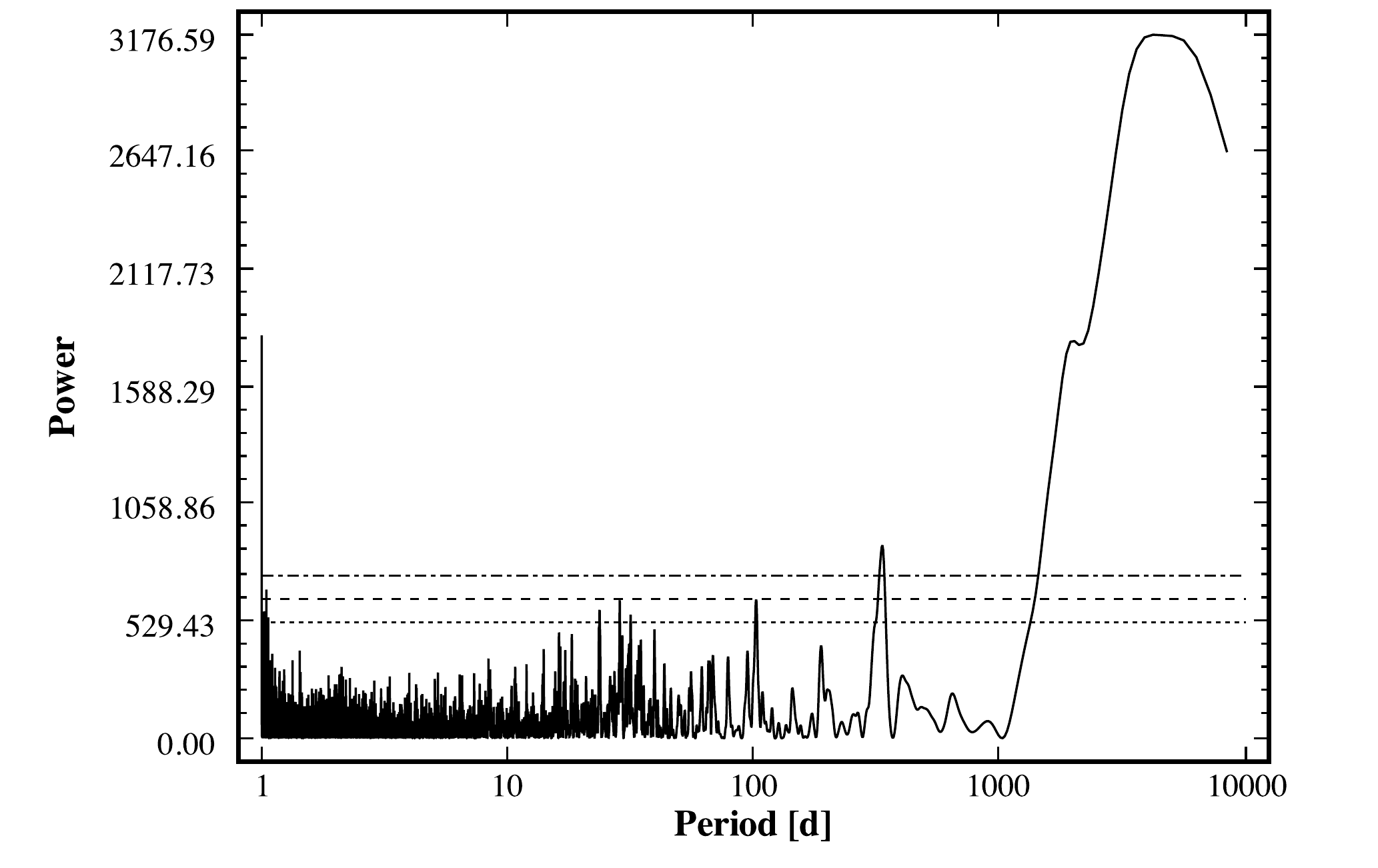}{0.49\textwidth}{}
          \fig{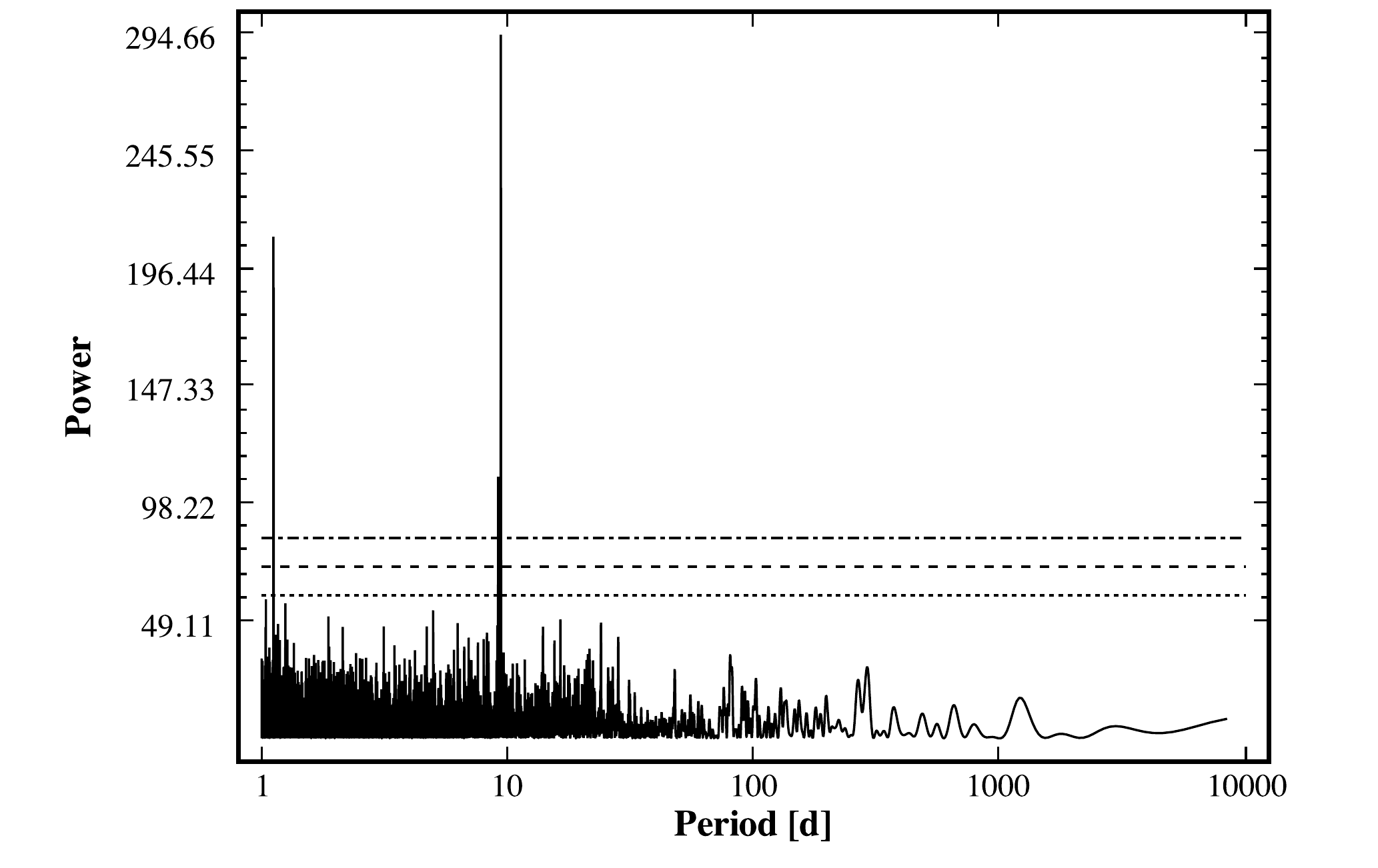}{0.49\textwidth}{}}
\caption{Left: Generalised Lomb-Scargle periodogram of the HD\,181433 
residuals after removal of the 1020-day planet.  A very long period 
signal is evident. Right: Generalised Lomb-Scargle periodogram of the HD\,181433 residuals after removing the 1014\,d and 7000\,d planets. Now the 9.37 day signal is apparent. Horizontal lines indicate false-alarm probabilities of 0.1\%, 1\%, and 10\%.}
\label{one}
\end{figure}

\begin{figure}
\gridline{\fig{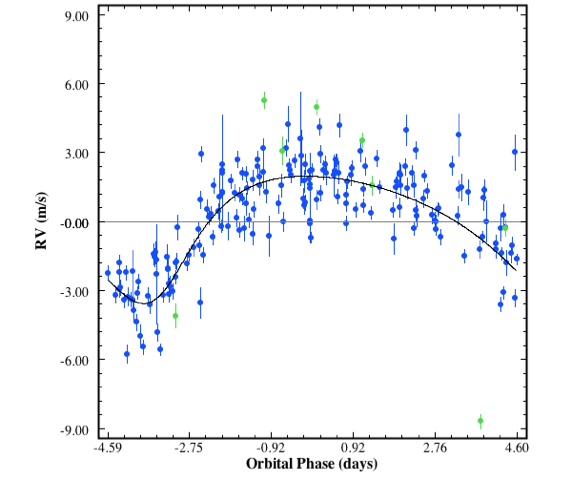}{0.33\textwidth}{}
          \fig{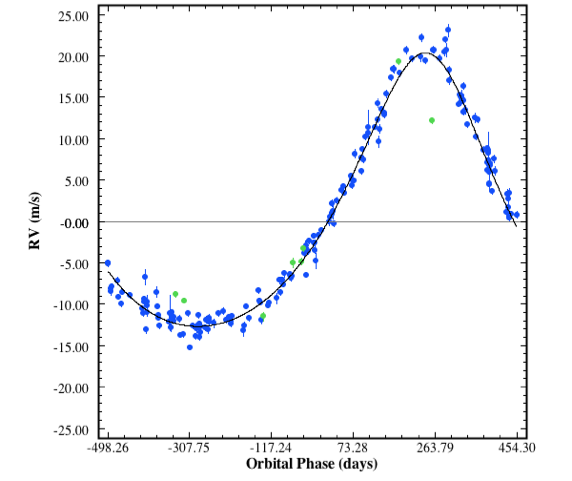}{0.33\textwidth}{}
          \fig{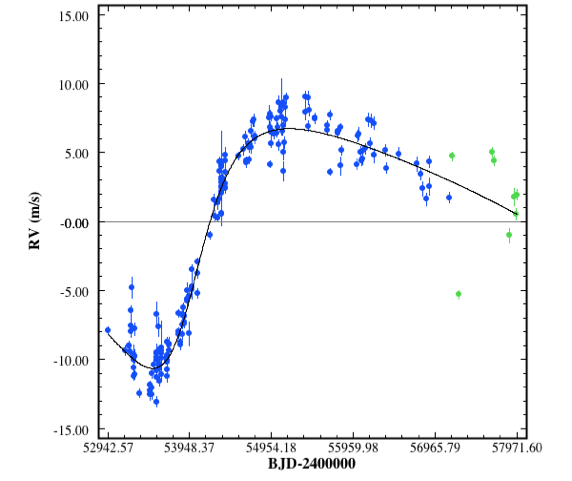}{0.33\textwidth}{}}
\caption{Data and model fit for the three planets in the HD\,181433 
system.  In each panel, the other two planets have been removed.  Left: 
planet b, centre: planet c, right: planet d. Here, the data plotted in green are those obtained after the HARPS fibre upgrade. Following that upgrade, HARPS began taking data again on May 19, 2015. The plots for the innermost planets (left two panels) show phase-folded data, for clarity.}
\label{3planetfit}
\end{figure}

In our new analysis, We used a total of 200 epochs, of which eight occurred after the 2015 May fibre upgrade and are treated as coming from a different instrument with its own velocity offset. Far and away the dominant signal in the periodogram is at $P\sim$1020 days; the left panel of Figure~\ref{one} shows the periodogram of the residuals after removing this planet.  The highest peak now lies at periods of several thousand days.   
The right-hand panel of Figure~\ref{one} shows the residuals periodogram after removing the second, long-period signal.  A clear and highly significant peak is now apparent at 9.37 days; our new analysis of the available HARPS data has so far recovered the three planets proposed in \citet{B09}, though with vastly different orbital parameters for the outer planet.  Notably, the eccentricity of the innermost planet fits best with $e_b=0.336\pm$0.014.  This is at first glance an improbably large eccentricity for a short-period planet, but \citet{B09} also arrived at a moderate $e=0.40\pm$0.06 for HD\,181433\,b.  We find zero-eccentricity solutions which are almost as good in a $\chi^2$ sense, but such orbits leave a residual signal of 4.68 days, exactly half the period of the innermost planet.  That strongly suggests that an eccentric orbit has been imperfectly removed; indeed it is the reverse of the situation we have encountered before \citep{shen08, ang10, 142paper, songhu}, where two circular planets can masquerade as a single eccentric planet.  We adopt the eccentric solution here and direct the interested reader to \citet{campanella13} for a discussion on the possible dynamical history of the system that could have produced such an orbit for planet b.  They propose that a previously ejected giant planet may have driven $e_b$ to its present value; alternatively, additional short-period low-mass planets could reproduce the observed system configuration.  At present, we do not see evidence for further short-period planets in this system.  The results of our three-planet fit are given in Table~\ref{3planets}, with uncertainties obtained from 10,000 bootstrap realisations within the \textit{Systemic Console}.  Data and model fits to the individual planetary signals are shown in Figure~\ref{3planetfit}. Our new fit has an rms of 1.39\,\ms, and has no significant residual signals.

%----------------------------------------------

\begin{deluxetable}{lrrr}
\tabletypesize{\scriptsize}
\tablecolumns{4}
\tablewidth{0pt}
\tablecaption{Adopted 3-Planet solution for HD\,181433}
\tablehead{
\colhead{Parameter} & \colhead{HD\,181433b} & \colhead{HD\,181433c} & \colhead{HD\,181433d}}
\startdata
\label{3planets}
Period (days) 			& 9.3745$\pm$0.0002 	& 1014.5$\pm$0.6 		& 7012$\pm$276 		\\
$T_0$ (BJD-2400000) 	& 52939.16$\pm$0.06 		& 52184.3$\pm$1.9 	& 46915$\pm$239 		\\ 
Eccentricity 			&  0.336$\pm$0.014 		& 0.235$\pm$0.003 	& 0.469$\pm$0.013 		\\
$\omega$ ($^\circ$) 		&  210.4$\pm$2.5 			& 8.6$\pm$0.7 		& 241.4$\pm$2.4 			\\
$K$ (\ms) 				&  2.7$\pm$0.1 		& 16.55$\pm$0.07 	& 8.7$\pm$0.1 		\\ 
m sin $i$ (\Mjup) 		&  0.0223$\pm$0.0003 		& 0.674$\pm$0.003 	& 0.612$\pm$0.004 		\\
$a$ (au) 				&  0.0801$\pm$0.0001 		& 1.819$\pm$0.001 	& 6.60$\pm$0.22 		\\
\enddata
\end{deluxetable}

%----------------------------------------------

\section{Dynamical Stability}\label{sec:NewDyn}

\begin{figure}
\includegraphics[width=1.0\textwidth]{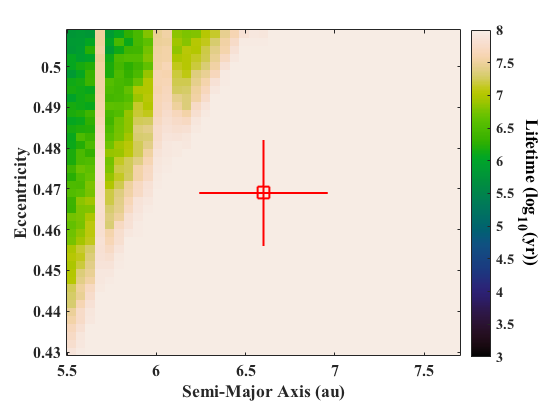}

\caption{The dynamical stability of the new three-planet solution for the HD~181433 system, as a function of the initial orbit of HD~181433~d. Despite the relatively large eccentricity of the orbit of HD~181433~d, the separation between the two outer planets in the system is such that our best-fit solution lies in a broad area of stability, far separated from the unstable region to the far left of the plot.}

\label{newdynamics}
\end{figure}

To assess the dynamical feasibility of our new three-planet solution for the 
HD\,181433, we performed two suites of dynamical simulations. The first 
were constructed in the same manner as the simulations described above. 
126,075 100 million year simulations were carried out, with an integration time-step of 40 days. In those simulations, the initial 
orbit of HD\,181433\,c being held fixed, and the orbit of HD\,181433\,d 
being varied across the $\pm3\sigma$ range in $a-e-\omega-M$ space. The 
result of these simulations describing the dynamical context of the 
orbit of HD\,181433\,d can be seen in Figure~\ref{newdynamics}. 

\begin{figure}
{\gridline{\fig{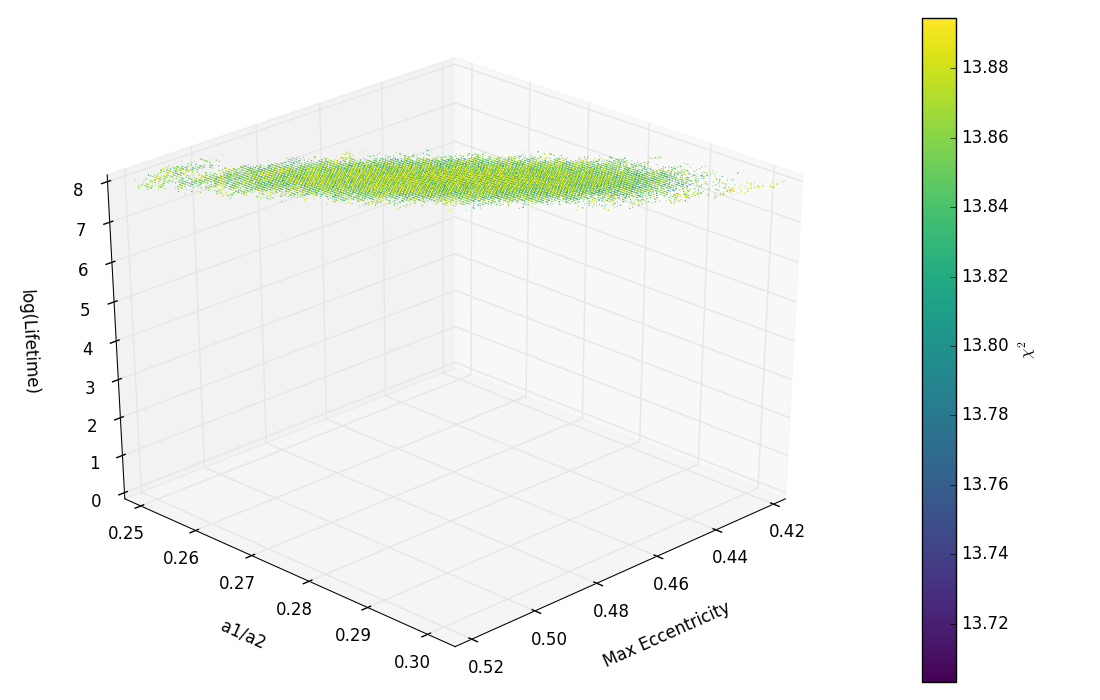}{0.49\textwidth}{}
          \fig{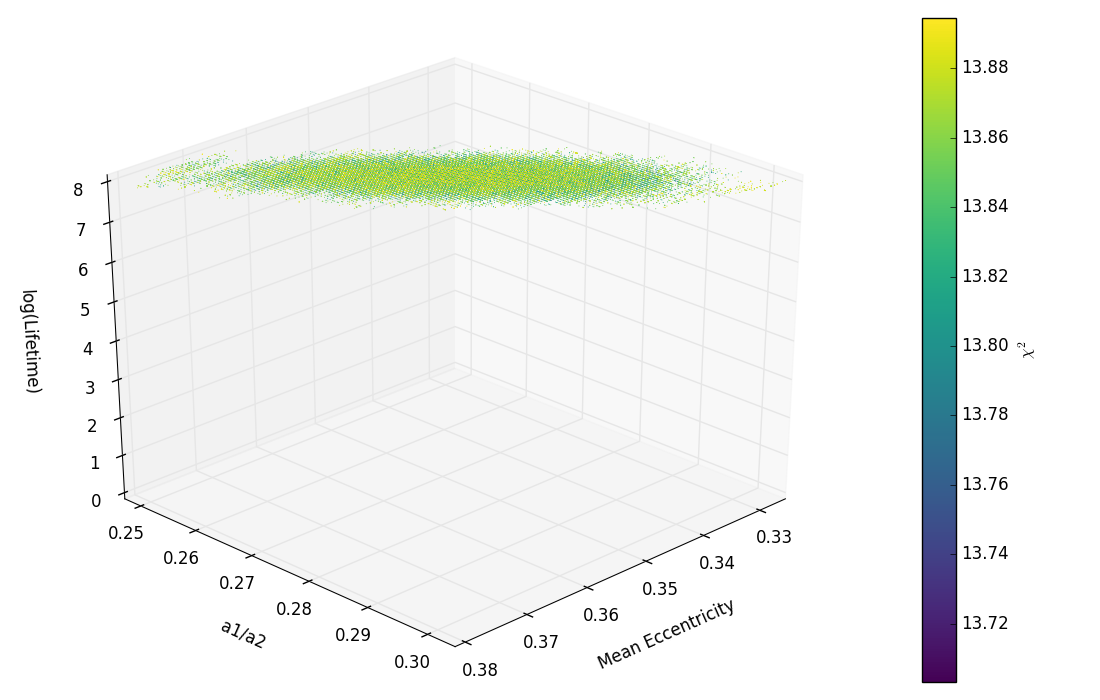}{0.49\textwidth}{}}
\gridline{\fig{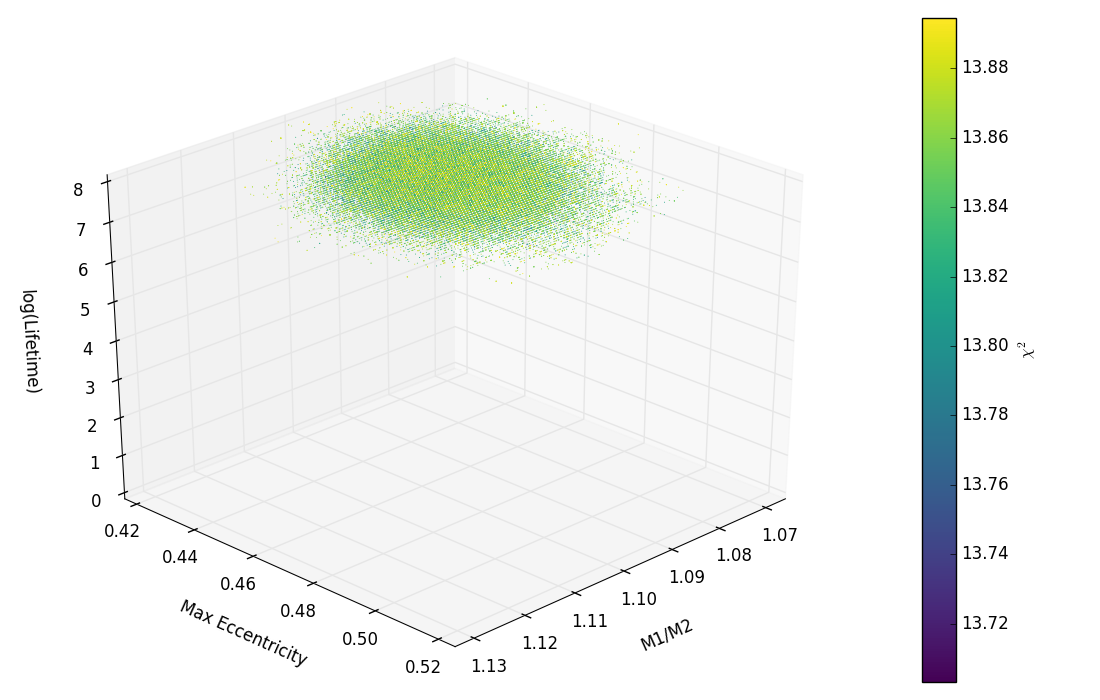}{0.49\textwidth}{}
          \fig{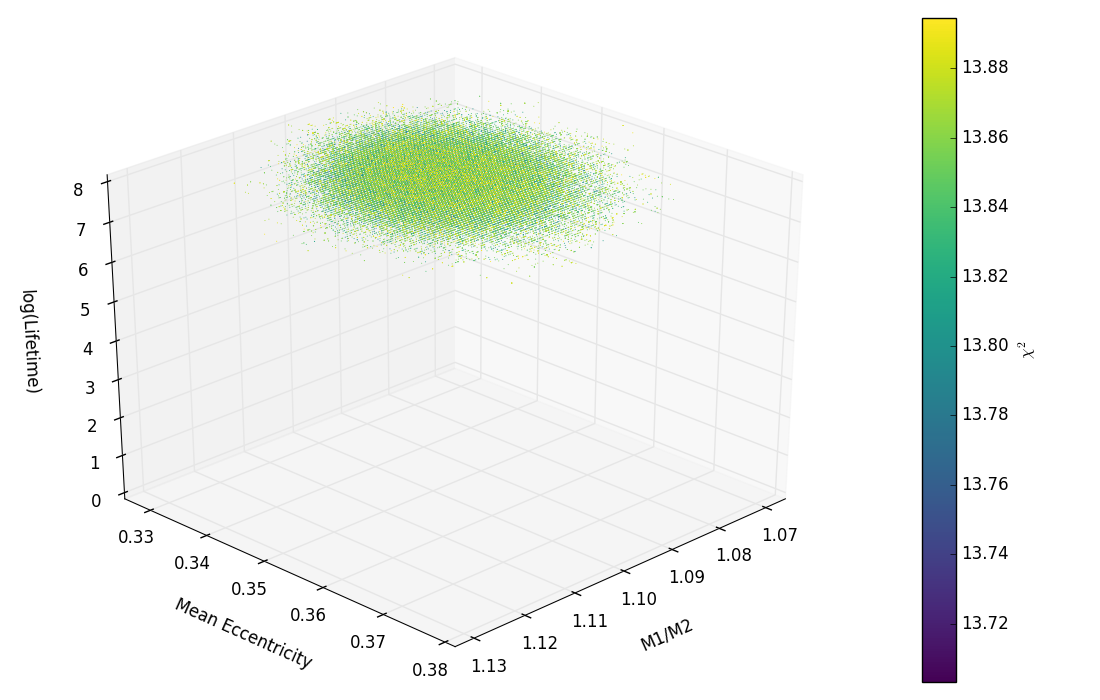}{0.49\textwidth}{}}
\gridline{\fig{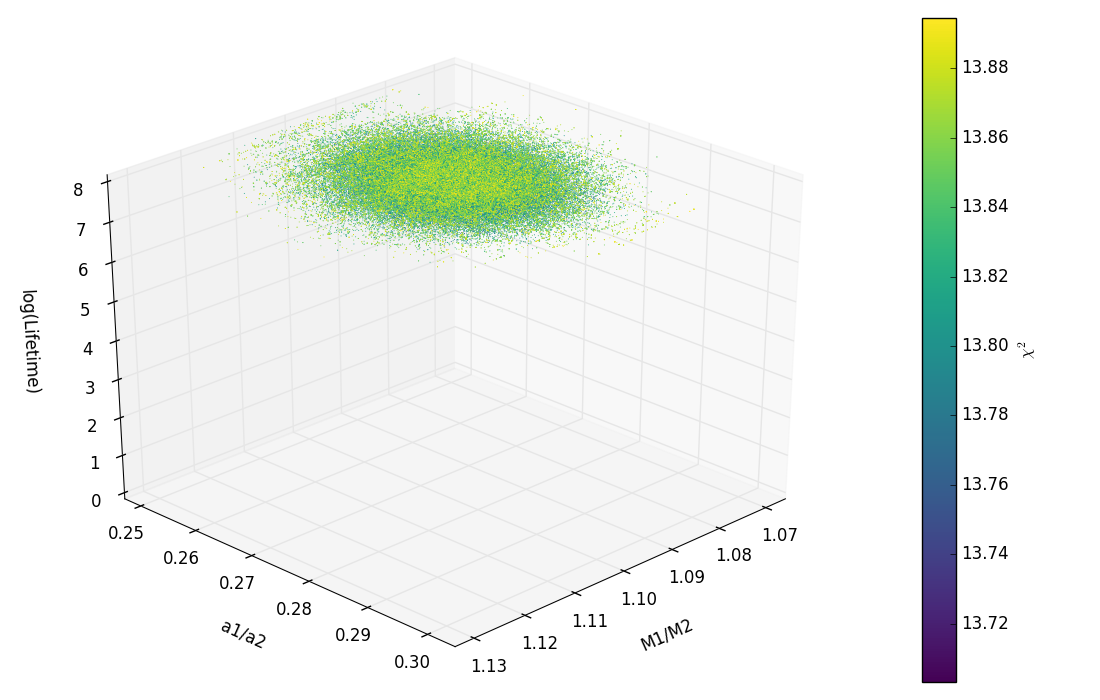}{0.49\textwidth}{}
          \fig{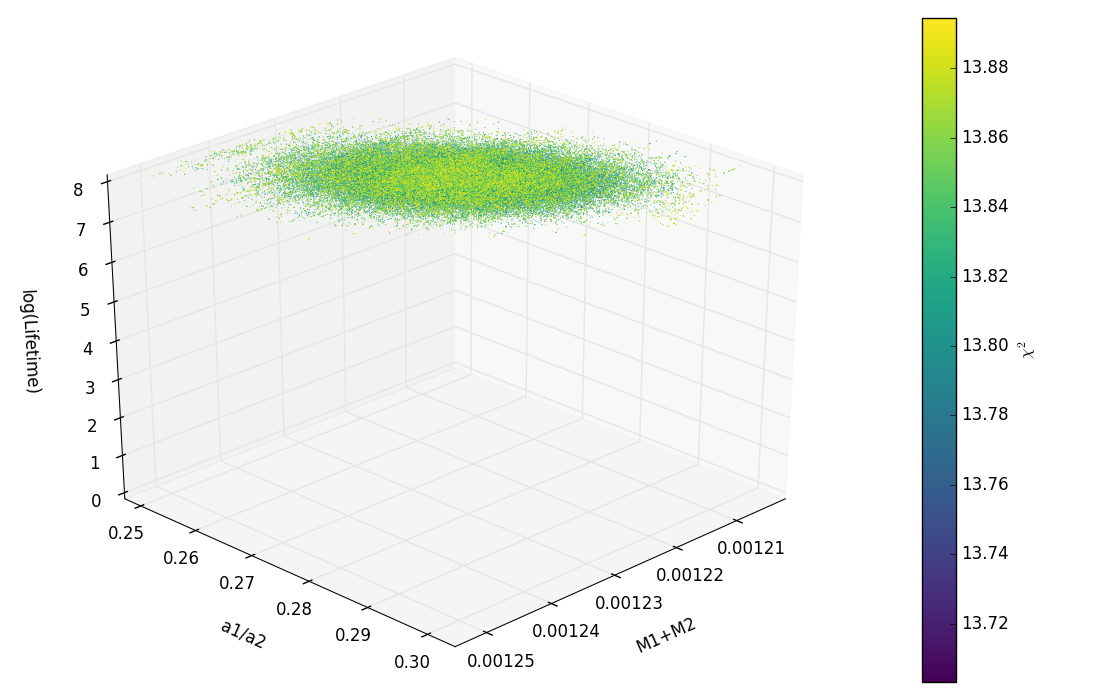}{0.49\textwidth}{}}
}
\caption{The dynamical stability of the new three-planet solution for the HD~181433 system, as a function of the initial orbital and physical parameters of the two outermost planets, c and d. These plots show the results of a cloud of simulations of planet-pairs, whose orbits and masses yield a good fit to the observational data. As can be seen, all 126,075 tested solutions proved to be dynamically stable for a period of 100~Myr (the full duration of the integrations). The distribution of points reveals the degree to which the various parameters are correlated, essentially mapping out the uncertainty ellipse in those parameter spaces around the best-fit solutions for the orbits of HD\,181433~c and d. }
\label{newmovies}
\end{figure}

We performed an additional suite of 126,075 simulations, 
following a new methodology first performed in our recent study of the 
newly discovered planetary system orbiting HD\,30177 \citep{30177}. 
Here, rather than simply move the orbit of one planet whilst keeping the 
other fixed, we instead generated 126,075 unique fits to the observational data, creating a cloud of solutions distributed around the surface in $\chi^2$ 
that covered the plausible solutions that fall within $\sim3\sigma$ of 
the best fit. The results of these simulations can be seen in 
Figure~\ref{newmovies}.

The results of those simulations showing the broader dynamical context of the new three-planet solution can be seen in Figure~\ref{newdynamics}. It is immediately clear that the new system architecture exhibits strong dynamical stability. Despite its eccentricity, the best-fit orbit for HD\,181433~d lies in a broad region of stability, with the only unstable region falling a significant distance from that solution in both semi-major axis and eccentricity space.

Figure~\ref{newmovies} presents the results of our simulations for the planetary pairs (HD\,181433~c-d) drawn from the 'clone cloud' distributed through the $\chi^2$ surface of plausible solutions for the system. The various sub-panels of that figure illustrate the way in which the different parameters of the fit are correlated to one another. Above all, however, they reveal that our new solution yields nothing but dynamically stable versions of the HD\,181433 system. Of 126,075 unique realisations tested in this manner, none exhibited dynamical instability on the 100~Myr timescale of our simulations, despite the eccentricity invoked for the outermost planet. 

%----------------------------------------------

%----------------------------------------------

\section{Discussion and Conclusions}  \label{sec:Discussion}

In this work, we have carried out a thorough and detailed study of the proposed planetary system orbiting the star HD\,181433. Our results show 
the critical importance of including studies of a system's dynamical feasibility in exoplanet discovery papers - particularly when the 
proposed solutions for the newly discovered planets feature moderate or 
high eccentricities, and/or large uncertainties. Such simulations can 
act as a `red flag,' revealing systems for which analysis of the 
observational data converges on solutions that are not dynamically 
feasible. For such systems, the results of dynamical simulations reveal 
the need for additional observational data, to help to better constrain 
the orbits of the planets suspected to lurk within.

In the case of HD\,181433's planetary system, we find that the orbital 
solution proposed in \citet{B09} is simply not dynamically feasible, 
unless the orbits of the outermost two planets (HD\,181433~c and d) move 
on orbits that are moderately inclined to one another, and also mutually 
resonant. The solution presented by \citet{C11}, by contrast, lies in a 
very narrow region of stability, engendered by mutual 5:2 mean-motion 
resonance between HD\,181433~c and d.

Since the publication of those two works, a significant number of new 
radial velocity observations have been made of the HD\,181433 system, 
and we therefore considered it prudent to fit that new data, to 
determine whether an improved solution was now available for the planets. 
In total, 200 radial velocities were used in our analysis, obtained from the publicly available HARPS spectra from the ESO archive. Using those 
data, we obtain a revised three-planet solution for the HD\,181433 system. That solution, presented in Table~\ref{3planets}, yields orbits for the innermost two planets in the system (HD~181433~b and c) that are very similar to those found by \citet{B09}. The best-fit orbit of HD\,181433~d, however, is changed markedly by the new data. Where the \citet{B09} solution placed that planet at $a = 3$~au, $e = 0.48$, and gave a mass of 0.54$M_{Jup}$, our new solution places it instead at $a \sim 6.6$ au, $e = 0.469$, with a mass of $\sim$0.612 $M_{Jup}$.

In stark contrast to the modified solution presented by \citet{C11}, we find that the 3-planet solution we propose for the HD\,181433 system is dynamically stable across a wide range of orbital parameter space. Indeed, when we integrated the orbits of 126,075 unique planet pairs (HD\,181433 c and d) drawn randomly from across the $3-\sigma$ uncertainty ellipse around the best fit orbit for a period of 100~Myr, every single tested pair proved dynamically stable (as can be seen in Figure~\ref{newmovies}), giving us confidence that the new solution is a fair representation of the true state of the HD\,181433 planetary system.

In light of the moderate eccentricities invoked by our fit for both HD\,181433~c and d, it is interesting to note that a number of recent studies \citep[e.g.][]{Rod09,142paper,songhu,kurster15,trifon17,witt19a,witt19b} have found that, under certain circumstances, two planets on low-to-moderately eccentric orbits can masquerade as a single highly eccentric planet in such fitting processes when data is sparse. As such, it might be natural to wonder whether the HD\`181433 system holds more surprises in the future -- and whether further observations might reveal the presence of additional planets in the system. We note, however, that, once the effects of the three planets proposed in this work have been removed from the data, we are left with no significant residual signals, and a low rms of just 1.39 ms$^{-1}$. As such, in this case, it seems that there is little need to invoke the presence of additional planets to explain the observed data.

Given the long period of the outermost planet proposed in this work (HD\,181433~d; 7012 days), we note that the temporal baseline covered by the observations of the system does not year fully encompass a single orbital period for that planet. As a result, we feel that future ongoing observations of this system are still needed in order to confirm the true nature of the outer planet, but the combination of the excellent fit our solution provides to the data, and the strong dynamical stability that that solution exhibits, provide confidence that the new solution is a fair reflection of the true nature of the system. In a broader sense, the HD\,181433 system stands as an important and illustrative `red flag,' highlighting the importance of undertaking a detailed dynamical analysis of newly discovered multi-planet systems, as a means to ensure that solutions presented to the wider community are feasible.

\acknowledgments

The work was supported by iVEC through the use of advanced computing resources located at the Murdoch University, in Western Australia. This 
research has made use of NASA's Astrophysics Data System (ADS). This research has made use of the SIMBAD database, operated at CDS, Strasbourg, France.
Part of the numerical simulations were performed by using the high performance computing cluster at the Korea Astronomy and Space Science Institute (KASI). 
TCH is supported by KASI research grant 2016-1-832-01 and 2017-1-830-0. JPM acknowledges this research has been supported by the Ministry of Science and Technology of Taiwan under grants MOST104-2628-M-001-004-MY3 and MOST107-2119-M-001-031-MY3, and Academia Sinica under grant AS-IA-106-M03. This work made use of publicly-available HARPS spectra from the ESO Archive, with the following program IDs: 072.C-0488(E), 192.C-0852(A), 198.C-0836(A), 183.C-0972(A), 077.C-0364(E), 091.C-0936(A), 60.A-0936(A). The authors appreciate the input and feedback from the anonymous referee, which resulted in an improved final manuscript.

\vspace{5mm}
\facilities{Katana (UNSW); Fawkes (USQ); iVEC (Murdoch University)}

\software{{\sc Mercury} \citep{Mercury}}



\section*{Appendix I: Radial Velocities used in this work}

%\begin{deluxetable}{lrr}
\begin{longtable}{lrr}
%\tabletypesize{\scriptsize}
%\tablecolumns{3}
%\tablewidth{0pt}
\caption{HARPS Radial Velocities for HD\,181433} \\
%\tablehead{
\textbf{BJD-2400000} & \textbf{RV (m/s)} & \textbf{Uncertainty} \\
\hline
%\startdata
\endhead
\label{rvdata}
52942.56654  &      -6.8  &    0.3  \\
53153.85493  &       7.7  &    0.4  \\
53202.69645  &      11.4  &    0.4  \\
53204.67449  &      13.4  &    0.3  \\
53217.71181  &      10.4  &    0.4  \\
53229.65203  &       7.7  &    0.4  \\
53230.68560  &      12.1  &    0.3  \\
53232.64333  &      13.3  &    0.8  \\
53237.73082  &      11.6  &    0.8  \\
53263.59448  &       5.6  &    0.3  \\
53265.56261  &       4.3  &    0.3  \\
53266.54601  &       2.2  &    0.4  \\
53267.55763  &       2.4  &    0.3  \\
53268.57529  &       5.7  &    0.3  \\
53269.57992  &       7.3  &    0.7  \\
53271.54520  &       6.9  &    0.5  \\
53272.55226  &       6.3  &    0.5  \\
53273.56639  &       4.0  &    0.3  \\
53274.54610  &       5.9  &    0.4  \\
53340.52541  &      -7.7  &    0.3  \\
53465.89979  &     -17.0  &    0.3  \\
53466.89181  &     -16.7  &    0.4  \\
53468.86848  &     -16.5  &    0.3  \\
53484.87672  &     -17.8  &    0.3  \\
53491.86912  &     -23.7  &    0.3  \\
53492.85771  &     -20.1  &    0.3  \\
53511.86498  &     -19.7  &    0.4  \\
53542.71667  &     -19.3  &    0.3  \\
53543.72206  &     -18.2  &    0.4  \\
53544.77493  &     -18.6  &    0.4  \\
53545.82160  &     -16.9  &    0.9  \\
53547.75035  &     -26.5  &    0.4  \\
53549.82048  &     -19.8  &    1.2  \\
53550.69452  &     -18.2  &    0.4  \\
53551.72523  &     -19.1  &    0.5  \\
53572.78012  &     -17.5  &    0.7  \\
53575.70397  &     -23.3  &    0.3  \\
53576.67957  &     -24.8  &    0.3  \\
53577.75843  &     -21.3  &    0.3  \\
53578.73270  &     -21.1  &    0.3  \\
53604.65670  &     -24.3  &    2.2  \\
53606.67931  &     -21.3  &    0.5  \\
53607.63178  &     -19.7  &    0.3  \\
53608.67638  &     -19.4  &    0.3  \\
53609.60901  &     -19.4  &    0.4  \\
53668.53352  &     -23.4  &    0.4  \\
53670.58735  &     -25.0  &    0.4  \\
53671.57257  &     -20.6  &    0.3  \\
53672.58635  &     -20.2  &    0.4  \\
53673.59598  &     -21.5  &    0.4  \\
53674.55888  &     -20.4  &    0.3  \\
53675.59932  &     -21.7  &    0.3  \\
53694.50090  &     -19.8  &    0.4  \\
53694.50476  &     -19.9  &    0.5  \\
53694.50854  &     -20.3  &    0.4  \\
53810.90026  &     -19.6  &    0.3  \\
53813.89414  &     -15.5  &    0.3  \\
53815.90144  &     -16.1  &    0.3  \\
53833.91339  &     -15.8  &    0.3  \\
53835.91447  &     -16.8  &    0.3  \\
53861.83976  &     -13.7  &    0.3  \\
53863.85258  &     -13.3  &    0.3  \\
53865.85270  &     -16.1  &    0.2  \\
53867.87460  &     -15.7  &    0.3  \\
53870.82281  &     -11.1  &    0.4  \\
53882.85095  &     -12.6  &    0.2  \\
53883.83516  &     -13.8  &    0.2  \\
53886.86707  &     -13.9  &    0.3  \\
53887.81637  &     -11.8  &    0.3  \\
53917.81053  &      -7.7  &    0.4  \\
53919.79936  &      -8.5  &    0.5  \\
53921.74104  &     -10.1  &    0.5  \\
53944.67109  &      -8.2  &    0.9  \\
53950.70411  &      -9.8  &    0.3  \\
53976.58348  &      -3.6  &    0.3  \\
53980.63197  &      -2.7  &    0.4  \\
53981.66303  &      -2.1  &    0.3  \\
53982.66112  &      -1.4  &    0.3  \\
53983.62520  &      -1.3  &    0.9  \\
54049.51371  &       5.0  &    0.3  \\
54051.53703  &       5.4  &    0.3  \\
54053.52956  &       3.4  &    0.3  \\
54199.88726  &      21.5  &    0.3  \\
54254.82740  &      21.7  &    0.4  \\
54255.75459  &      20.3  &    0.4  \\
54291.79520  &      17.5  &    0.5  \\
54296.78082  &      12.3  &    0.3  \\
54314.73552  &      14.7  &    0.5  \\
54316.59151  &       9.3  &    0.4  \\
54320.77830  &      16.8  &    0.3  \\
54342.65548  &      11.6  &    0.4  \\
54343.72407  &       8.3  &    0.5  \\
54344.70570  &       6.6  &    0.4  \\
54345.68419  &       8.2  &    0.3  \\
54346.70956  &      11.9  &    2.5  \\
54346.71180  &      11.0  &    1.5  \\
54346.71402  &      11.0  &    1.4  \\
54346.71618  &      12.0  &    1.0  \\
54346.71844  &      12.3  &    1.1  \\
54348.67547  &       9.5  &    0.8  \\
54348.67766  &       9.6  &    0.9  \\
54348.67992  &      11.7  &    0.8  \\
54348.68220  &      11.2  &    0.8  \\
54348.68434  &      11.0  &    0.8  \\
54349.61971  &      11.8  &    0.3  \\
54350.63331  &      11.1  &    0.3  \\
54388.58439  &       5.6  &    0.5  \\
54389.58709  &       6.7  &    0.6  \\
54391.56659  &       2.9  &    0.9  \\
54392.54769  &       3.1  &    0.2  \\
54393.56845  &       4.6  &    0.3  \\
54394.57102  &       6.0  &    0.5  \\
54554.89327  &      -3.2  &    0.3  \\
54616.93561  &      -9.9  &    0.4  \\
54639.89084  &      -3.9  &    0.4  \\
54642.75960  &      -8.0  &    0.3  \\
54648.54964  &      -5.7  &    0.3  \\
54672.76897  &      -9.9  &    0.3  \\
54677.81501  &      -6.0  &    0.3  \\
54681.79082  &      -9.7  &    0.4  \\
54700.73214  &      -9.8  &    0.3  \\
54703.72717  &      -3.9  &    0.4  \\
54707.71419  &      -6.3  &    0.4  \\
54732.49341  &      -2.7  &    0.4  \\
54743.55239  &      -2.7  &    0.4  \\
54749.52266  &      -5.4  &    0.3  \\
54759.56437  &      -4.2  &    0.3  \\
54933.86536  &       2.8  &    0.9  \\
54935.91833  &      -2.7  &    0.2  \\
54939.91893  &       5.3  &    0.4  \\
54941.85017  &       5.4  &    0.2  \\
54953.89384  &       2.1  &    0.3  \\
54954.84654  &       1.7  &    0.2  \\
54955.83353  &       3.8  &    0.3  \\
54956.87791  &       6.1  &    0.4  \\
54988.87407  &       7.6  &    0.4  \\
55021.87307  &      12.3  &    0.3  \\
55024.83196  &      12.2  &    0.3  \\
55040.70437  &      13.8  &    0.4  \\
55041.69051  &      13.4  &    0.4  \\
55048.77824  &      13.2  &    0.5  \\
55072.64646  &      18.5  &    0.4  \\
55079.68145  &      20.5  &    2.0  \\
55079.70027  &      19.8  &    1.2  \\
55095.62794  &      16.4  &    0.3  \\
55102.54122  &      20.6  &    0.5  \\
55105.55924  &      16.4  &    0.7  \\
55106.54410  &      19.7  &    0.5  \\
55110.55571  &      21.4  &    0.4  \\
55117.54270  &      22.2  &    0.4  \\
55123.57348  &      19.8  &    0.5  \\
55134.50271  &      25.7  &    0.4  \\
55138.51140  &      26.5  &    0.4  \\
55373.69437  &      14.7  &    0.4  \\
55376.65744  &      10.4  &    0.5  \\
55408.65360  &      12.4  &    0.5  \\
55413.72073  &       4.4  &    0.4  \\
55425.68836  &       9.6  &    0.4  \\
55487.53307  &      -0.0  &    0.4  \\
55488.50616  &      -1.7  &    0.3  \\
55640.91832  &      -3.4  &    0.3  \\
55642.91749  &      -3.3  &    0.3  \\
55674.86893  &      -6.1  &    0.3  \\
55679.87111  &      -6.6  &    0.2  \\
55769.72708  &      -8.5  &    0.3  \\
55777.76528  &      -6.3  &    0.4  \\
55803.60759  &      -5.4  &    0.7  \\
55809.57178  &      -2.4  &    0.3  \\
55816.56622  &      -8.7  &    0.3  \\
56013.89530  &       2.5  &    0.4  \\
56021.85348  &       6.7  &    0.4  \\
56032.87656  &       7.1  &    0.4  \\
56056.83457  &      12.2  &    0.4  \\
56061.80773  &      10.7  &    0.3  \\
56079.76887  &      10.9  &    0.3  \\
56082.80046  &      15.4  &    0.4  \\
56117.82953  &      17.3  &    0.4  \\
56154.48256  &      20.9  &    0.5  \\
56167.61926  &      25.6  &    0.4  \\
56182.54979  &      23.1  &    0.5  \\
56216.56084  &      26.5  &    0.5  \\
56218.53608  &      26.3  &    0.5  \\
56362.89849  &      15.6  &    0.4  \\
56371.89839  &      12.7  &    0.4  \\
56525.72207  &      -1.1  &    0.5  \\
56748.89833  &      -5.9  &    0.5  \\
56789.84570  &      -8.7  &    0.4  \\
56819.72394  &     -11.9  &    0.5  \\
56859.74068  &      -6.2  &    0.5  \\
56896.68485  &      -4.3  &    0.6  \\
56903.60307  &      -6.0  &    0.4  \\
57146.86582  &      14.2  &    0.4  \\
\hline
57180.93319  &      29.4  &    0.3  \\
57258.57711  &      19.6  &    0.3  \\
57675.56374  &      -0.1  &    0.3  \\
57695.49173  &      -0.7  &    0.3  \\
57879.80718  &      -7.4  &    0.5  \\
57947.84004  &       3.0  &    0.6  \\
57968.59812  &       2.8  &    0.4  \\
57971.59547  &       1.1  &    0.4  \\
%\enddata
\end{longtable}
%\end{deluxetable}

\end{document}